\documentclass[aps,prd,showpacs,nofootinbib,11pt]{revtex4-1}
\usepackage{amsmath,amsfonts,amssymb,graphicx,accents,mathrsfs}
\usepackage{cleveref,braket,textcomp}
\usepackage[percent]{overpic}
\usepackage{color}
\DeclareFontFamily{U}{rsfs}{}         
\DeclareFontShape{U}{rsfs}{m}{n}{<5> rsfs5 <6><7> rsfs7          %
  <8><9><10><10.95><12><14.4><17.28><20.74><24.88> rsfs10}{}     %
\DeclareMathAlphabet{\mathfs}{U}{rsfs}{m}{n}                     %
\def\beq{\begin{eqnarray}}
\def\eeq{\end{eqnarray}}

\def\nn{\nonumber\\}

\DeclareMathOperator{\sech}{sech}

\begin{document}
\title{\boldmath Generalized geodesic deviation equations and entanglement first law for rotating BTZ black holes}
\author{Avirup Ghosh}\email{avirup.ghosh@saha.ac.in}
\affiliation{Theory Division, Saha Institute of Nuclear Physics, 1/AF Bidhan 
Nagar, Kolkata 700064, INDIA.}
\author{Rohit Mishra,}\email{rohit.mishra@saha.ac.in}
\affiliation{Theory Division, Saha Institute of Nuclear Physics, HBNI, 1/AF Bidhan 
Nagar, Kolkata 700064, INDIA}

\begin{abstract}
The change in Holographic entanglement entropy (HEE) for small fluctuations about pure AdS is given by a perturbative expansion of the  area functional in terms of the change in the 
bulk metric and the embedded extremal surface. However it is known that change in the embedding appear  at second order or higher. In this paper we show that these changes in the 
embedding can be systematically calculated in the 2+1 dimensional case by accounting for the deviation of the spacelike geodesics between a spacetime and perturbations over it. 
Here we consider rotating BTZ as perturbation over $AdS_3$  and study deviations of spacelike geodesics in them. We argue that these deviations arise naturally as solutions of 
a ``generalized geodesic deviation equation''. Using this we perturbatively calculate the changes in HEE upto second order, for rotating BTZ. This expression matches with the  
small system size expansion of the change in HEE obtained by HRT (Hubeny, Rangamani and Takayanagi) proposal for rotating BTZ. We also write an alternative form of entanglement first law 
for rotating BTZ. To do this one needs to go beyond the leading order in the perturbation series discussed above. That's precisely the reason we consider finding a systematic way to calculate it. To put our result on a firm footing we further show that it is this alternative first law that approaches the thermal first law in the large subsystem size limit.
\end{abstract}

\maketitle
\section{Introduction}
The AdS/CFT correspondence \cite{Malda,Gubser,Witten} has been a very successful idea in string theory. It relates conformal field 
theories living on the boundary of asymptotically anti de Sitter (AdS) spacetimes with the supergravity theory in the bulk. More 
precisely certain correlation functions in the boundary CFT can be obtained by calculating certain geometrical quantities in the bulk. 
Entanglement entropy has been very useful in studying correlation between nonlocal operators in quantum many body systems. In  the 
framework of AdS/CFT a proposal for calculating the entanglement entropy for subsystems in a strongly coupled quantum system living on 
the boundary of asymptotically $AdS$ spacetimes was given in \cite{Ryu1,Ryu2}. In this proposal the holographic entanglement entropy (HEE)
$S_A$ of a subsystem $A$ in the boundary $CFT$ is given by the area of a codimension $2$ minimal surface ${\gamma_A}$ in bulk viz.
\begin{gather}\label{HEE}
 S_A ={Area(\gamma_A)\over 4 G_N}.
\end{gather}
Where $G_N$ is the Newton's constant
For cases where the bulk metric is static, 
let $\frac{\partial}{\partial t}$ be the timelike, hypersurface orthogonal Killing vector in the bulk. Then $\gamma_A$ is to be 
understood as the area of the minimal co dimension two surface, homologous to the boundary subsystem, on a $t=constant$ slice. When the bulk metric is not static or may be 
even non-stationary one must use the covariant HEE prescription (HRT proposal) proposed in \cite{Hubeny}. In this case $\gamma_A$ is to be understood as the 
area of a spacelike extremal surface in the bulk spacetime anchored to the boundary subsystem. 

Having obtained the holographic entanglement entropy for pure $AdS_{d+1}$ and  for a metric which is an excitation over pure $AdS$, one can 
find the difference of the entanglement entropy $\Delta S_E=S_{exc.}-S_{AdS_{d+1}}$. For small subsystems one can perturbatively expand 
$\Delta S_E$ order by order in some small dimensionless parameter related to subsystem size. Upto first order, in the small 
subsystem size approximation, it has been shown \cite{JT} that  the change in HEE satisfies a first law like relation in $CFT_d$, with a universal entanglement temperature $T_E$. 
If $\Delta E$ is the energy of the excitations for the subsystem above ground state, then first law is of the form
\begin{gather}
 \Delta S_E= {1\over T_E}\Delta E.
\end{gather}
For IR deformations (excitations) in asymptotically $AdS$ which carry additional charges like angular momentum, pressure etc., the form of entanglement first law gets modified \cite{alisha}. Similar expressions have been obtained for non conformal and higher derivative gravity backgrounds in \cite{,Park,Pang,He:2013rsa,Guo:2013aca}. 
Recently it has been shown \cite{He,Kim:2015rvu,Blanco,mishra} that the first law like relation also holds if one considers perturbations upto second order. 

For $AdS_{d+1}$ the minimal surface $\gamma_A$ are $(d-1)$ dimensional. The surface $\gamma_A$ is an extremum of the area functional
\begin{gather}
 Area=\int d^{d-1}\sigma\sqrt{h},
\end{gather}
where $\sigma$ are the coordinates and $h_{ab}$ is the induced metric on $\gamma_A$. Variation of the area functional depends both on metric perturbations and variation of the minimal surface itself.
Change in HEE at each order can be obtained by subtracting the pure $AdS$ contribution from the variation of the area functional. At first order, contributions from changes in the shape of the extremal surface does not appear
as $\gamma_A$ satisfies extremal condition on the background $AdS$ geometry \cite{Lashkari,Jyotirmoy,Nozaki}. 

In this paper we propose a way to calculate second order variations of  the area functional by taking
into account changes in both metric perturbation and shape of the extremal surface in $2+1$ dimension. This is achieved by studying geodesic deviations between geodesics in rotating BTZ black hole ( seen as perturbation over pure $AdS$) and $AdS_3$. As will be clear from the construction these deviations can be obtained as solutions of a ``generalized geodesic deviation equation'' (\cite{Pyne:1993np} and references therein).
Second order expressions for HEE obtained from variation of the area functional matches exactly with the second order expansion of HEE obtained by HRT proposal. We also present an alternative form of first law
of entanglement thermodynamics which involves the differential change in $\Delta S~(d\Delta S~\text{for example})$ rather than $\Delta S$ itself. The modified first law includes contributions from angular momentum of the $BTZ$ backgound and approaches the first law of black hole thermodynamics in large $l$ (the subsystem size) limit. 

It turns out that
2+1 dimensional gravity has no propagating degrees of freedom and therefore exact analytical expressions for certain quantities can be found. This is precisely the reason why 2+1 dimensional gravity can be written as  Chern Simons theory. Chern Simons is topological and therefore the solutions depend only on the topology of the underlying manifold. From a geometric stand point the Weyl tensor identically vanishes in 3 dimensions and therefore the Riemann tensor is completely specified by the Ricci. As a consequence the solutions of
pure gravity with negative cosmological constant are necessarily locally isometric to pure $AdS_3$.
Despite this, there is still a rich set of solutions that differ from $AdS$ globally. 
One such solution is the rotating
BTZ black hole \cite{Banados:1992wn}. Since in the $2+1$ dimensional case exact expressions for the change in entanglement entropy and the minimal surface in $BTZ$ is known, one might question the need of a perturabative analysis. But it turns out that the $2+1$ dimensional case is a perfect ground for checking such proposals. It is to be noted that for higher dimensional case the change in entanglement entropy might have to be calculated pertubatively and hence a precise prescription is required. We should point out that the notion of ``deviations" of codimension two surfaces is well known for higher dimensions. Though algebraically difficult it is absolutely possible to find ``generalized deviation equations" for codimension two surfaces for dimensions $4$ or higher.
\section*{Plan of the paper}
In the section \ref{GDE} we give a brief derivation of the ``generalized deviation equations''. Though this is  given in 
\cite{Pyne:1993np}, we account for certain generalizations over the existing form and make the construction more precise from the point of view of covariant perturabations. The solutions of this gives a deviation between spacelike geodesics in, $AdS_3$ and a perturbations over it. In section \ref{VOL} we derive an expression for the variation of the
length functional upto second order perturbations of the metric and first order perturbations of the embedding. In section \ref{BTZgeo} we solve the generalized deviation equations for rotating BTZ like perturbations over $AdS_3$. These solutions are substituted in the expression for variation of length to get an 
expression for the change in HEE (upto second order) in terms of the components of the holographic stress tensor. In section \ref{FL} the expression thus obtained is verified with the expression for change in HEE (upto second order) for rotating BTZ obtained using HRT proposal. A generalized entanglement first law (upto second order) is also obtained. Which includes contributions from angular momentum.
\section{The generalized gedesic deviation equations}\label{GDE}
The generalized deviation equations have been known for quite some time, applications of which in the case of perturbed cosmological
spacetimes can be found in \cite{Pyne:1993np} (and references therein) where perturbed null geodesics are studied for perturbations 
around Einstein-de Sitter universe. In the holographic context generalized deviations of null geodesics in $AdS_3$ has been used, only recently, in \cite{Engelhardt:2016aoo} however in a 
very different context from ours. In the holographic entanglement entropy context codimension two minimal surfaces in $2+1$ dimemsions are spacelike geodesics. It is clear that the spacelike geodesics in $AdS_3$
anchored to the boundary subsystem are perturbed as one considers excitations over $AdS_3$. If we consider the variation of the area functional (HEE) 
$A(G,X^\mu)=\int \sqrt{\det{h}}~d^n\sigma, ~where ~h_{ab}=g_{\mu\nu}\frac{\partial X^\mu}{\partial\sigma^a}\frac{\partial X^\nu}{\partial\sigma^b}$, 
The variation of the quantity is therefore, 
\beq
A(G+\delta G,X^\mu+\delta X^\mu)-A(G,X^\mu),
\eeq
where $\delta X^\mu$ is the change of the embedding functions. To first order $\delta X^\mu$ does not contribute. The $\delta X^\mu$ starts contributing only at second order. Therefore while considering second order variations, perturbations of spacelike geodesics also contribute. These changes in the embedding of the spacelike geodesics can be obtained by studying geodesic deviation between $AdS_3$ and the perturbed spacetime. To do this we use the following formulation.

\subsection{First order generalized deviations}
\begin{figure}[h]
\caption{The mapping of the geodesics}
\includegraphics[width=20cm]{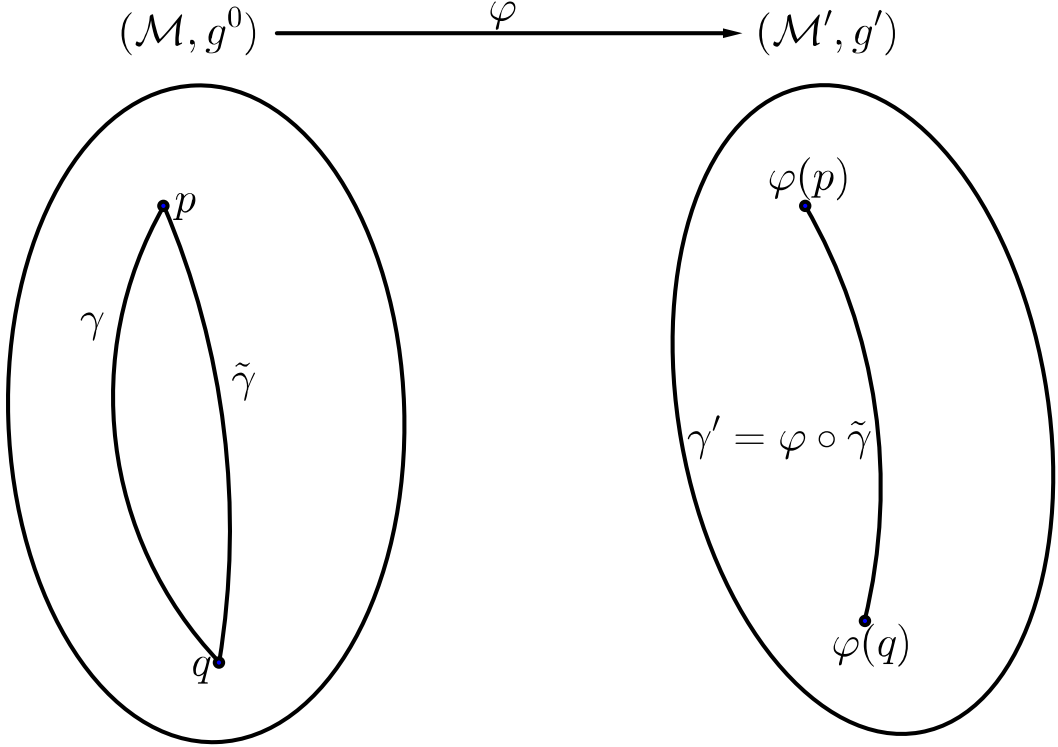}
\label{fig.geo}
\centering
\end{figure}

Consider an affinely parametrized geodesic parametrized by $\tau$ in a spacetime $(\mathcal M,\accentset{0}{g})$ with end points $p,q~\epsilon~\mathcal M$. 
\beq
\frac{d^2x^\mu}{d\tau^2}+\Gamma^{\mu}_{\nu\rho}(x)\frac{dx^\nu}{d\tau}\frac{dx^\rho}{d\tau}=0,
\eeq
where $\accentset{0}{\Gamma}^\mu_{\nu\rho}$ are the Christoffel symbols on $\mathcal M$ compatible with $\accentset{0}{g}$. Consider another spacetime $(\mathcal M',g')$. $(\mathcal M',g')$ is said to be a perturbation over $(\mathcal M,\accentset{0}{g})$ if there exists a diffeomorphism $\varphi:\mathcal M\rightarrow \mathcal M'$ such that $\varphi_{*}g'-\accentset{0}{g}=\accentset{(1)}{h}$ is a small perturbation over the unperturbed metric $\accentset{0}g$. 
Let $\gamma'$ be a geodesic in $\mathcal M'$ with parameter $\tau$ and end points $\varphi(p),\varphi(q)~\epsilon~\mathcal M'$. However it may not be affinely parametrized by $\tau$. Let $\tilde\gamma$ be a curve in $\mathcal M$ such that $\varphi\circ\tilde\gamma=\gamma'$. Therefore the tangent vector to $\gamma'$ in $\mathcal M'$ is essentially the push forward of the tangent vector of $\tilde\gamma$ in $\mathcal M$ (fig. \ref{fig.geo}). 
On $\mathcal M$, therefore $\tilde \gamma$ must satisfy,
\beq
\frac{d^2\tilde x^\mu}{d\tau^2}+\tilde\Gamma^{\mu}_{\nu\rho}(\tilde x)\frac{d\tilde x^\nu}{d\tau}\frac{d\tilde x^\rho}{d\tau}=f(\tilde x)\frac{d\tilde x^\mu}{d\tau},
\eeq
where $\tilde\Gamma$ are the Christoffels symbols on $\mathcal M$ compatible with $\varphi_{*}g'$. Note that $\tilde\gamma$ is not geodesic in $\mathcal M$ with respect to the initial Christoffels $\Gamma$. Let us assume that $\tilde\gamma$ is a small deviation about the curve $\gamma$. Therefore to first order we can write $\tilde x^{\mu}(\tau)=x^{\mu}(\tau)+\accentset{(1)}{\eta}^\mu(\tau)$. 
We also note that to first order in metric perturbations,
\begin{gather}
\tilde\Gamma^{\mu}_{\nu\rho}(x)=\accentset{0}{\Gamma}^{\mu}_{\nu\rho}(x)+\frac{1}{2}\accentset{0}{g}^{\mu\sigma}\left(\partial_\nu \accentset{(1)}{h}_{\rho\sigma}+\partial_\rho \accentset{(1)}{h}_{\nu \sigma}-\partial_\sigma \accentset{(1)}{h}_{\nu \rho}\right)-\frac{1}{2}\accentset{(1)}{h}^{\mu \sigma}\left(\partial_\nu\accentset{0}{g}_{\rho \sigma}+\partial_\rho\accentset{0}{g}_{\nu \sigma}-\partial_\sigma\accentset{0}{g}_{\nu \rho}\right)\nn
=\accentset{0}{\Gamma}^{\mu}_{\nu\rho}(x)+\accentset{(1)}{C}^{\mu}_{\nu \rho}(x)
\end{gather}
Therefore to first order,
\beq
\tilde\Gamma^{\mu}_{\nu\rho}(\tilde x)=\accentset{0}{\Gamma}^{\mu}_{\nu\rho}(x)+\accentset{(1)}{C}^{\mu}_{\nu \rho}(x)+\partial_\sigma \accentset{0}{\Gamma}^{\mu}_{\nu\rho}(x)\accentset{(1)}{\eta}^\sigma
\eeq
Subtracting the two geodesic equations give,
\beq
\frac{d^2\accentset{(1)}{\eta}^{\mu}}{d\tau^2}+\partial_\sigma\accentset{0}{\Gamma}^{\mu}_{\nu\rho}\accentset{(1)}{\eta}^{\sigma}\frac{dx^\nu}{d\tau}\frac{dx^\rho}{d\tau}+\accentset{0}{\Gamma}^{\mu}_{\nu\rho}\frac{dx^\rho}{d\tau}\frac{d\accentset{(1)}{\eta}^{\nu}}{d\tau}+\accentset{0}{\Gamma}^{\mu}_{\nu\rho}\frac{dx^\nu}{d\tau}\frac{d\accentset{(1)}{\eta}^{\rho}}{d\tau}=-\accentset{(1)}{C}^{\mu}_{\nu\rho}(x)\frac{dx^\nu}{d\tau}\frac{dx^\rho}{d\tau}+\partial_\sigma f(x)\accentset{(1)}{\eta}^\sigma\frac{d x^\mu}{d\tau}
\eeq
and $f\bigg|_\gamma=0$, which essentially means that the initial curve is affinely parametrized. The left hand side can now be identified as just the left hand side of the Jacobi equation. Therefore,
\begin{equation}\label{IJE}
\frac{\mathcal D^2\accentset{(1)}{\eta}^{\mu}}{d\tau^2}+\mathcal R^{\mu}~_{\nu\rho\sigma}\frac{dx^\nu}{d\tau}\frac{dx^\sigma}{d\tau}\accentset{(1)}{\eta}^{\rho}=-\accentset{(1)}{C}^{\mu}_{\nu\rho}(x)\frac{dx^\nu}{d\tau}\frac{dx^\rho}{d\tau}+\partial_\sigma f(x)\accentset{(1)}\eta^\sigma\frac{d x^\mu}{d\tau}=F^\mu+\partial_\sigma f(x)\accentset{(1)}{\eta}^\sigma\frac{d x^\mu}{d\tau}\nn
\end{equation}
Where ${D\over d\tau}$ is the covariant derivative along $\gamma$ and $\mathcal R^{\mu}~_{\nu\rho\sigma}$ is the Riemann tensor wrt $\accentset{0}{g}$. Therefore the resulting equation is an inhomogeneous deviation equation. Note that if $\accentset{(1)}C$ is set the zero the resulting equation is just the deviation equation for a non-affinely parametrized congruence of geodesics in a given space-time (no metric perturbations). To solve this equation the best procedure is to consider a local basis $e^{\mu}_1$ which is parallely propagated along the initial geodesic and 
writing the deviation vector and the inhomogeneous terms, in terms of the local basis i.e $\eta^{(1)\mu}=\accentset{(1)}{\eta}^{A}e^\mu_A$ and $F^\mu=F^A e^{\mu}_{A}$. For $AdS_3$ background which is maximally symmetric $\mathcal R_{\mu\nu\rho\lambda}=-(g_{\mu\rho}g_{\nu\lambda}-g_{\nu\rho}g_{\mu\lambda})$\footnote{We have set the radius of the $AdS$ space  to ``one" here and in all subsequent calculations.} and therefore for space-like geodesics, the equation reduces to,
\begin{gather}
\frac{d^2\accentset{(1)}{\eta}^{A}}{d\tau^2}-\accentset{(1)}{\eta}^{A}=F^A ~~~~~~~~~~\text{for A=0,2}\\
\frac{d^2\accentset{(1)}{\eta}^{A}}{d\tau^2}-\partial_Bf(x)\accentset{(1)}\eta^B=F^A ~~~~~~~~~~~~~~~~\text{for A=1, B= 0 to 2},
\end{gather}
where $e^{\mu}_1=T^\mu$ is the tangent vector to the geodesic and $ \partial_A=e_A^\mu\partial_\mu$. Note that the non affinity term enters only the equation for the component of the deviation vector in the direction of the tangent vector. The equations for $A=0,2$ can obviously be solved and the resulting solutions can be put in the equation for $A=1$ to get a ordinary differential equation for $\eta^1$. However the equation for $\eta^1$ cannot be solved due to presence of the unknown function $f(x)$. But we will see that we actually won't be requiring a solution for $\eta^1$ for calculation of the variation of geodesic length. Note that unlike the original deviation equation where the component of the deviation vector in the direction of the geodesic can be set equal to zero, one may not be able to do the same here due to the inhomogenous term. More precisely, the deviation along a geodesic is pure gauge and can be removed by a reparametrization of the geodesic. That is to say that the deviations along the geodesic does not affect the length of the perturbed geodesic only if the perturbed curve is a geodesic of the same space-time $(\mathcal M, g)$. However since in our case the perturbed curve is a geodesic in some perturbed space-time $(\mathcal M, g')$ these might  actually become physical. But this does not happen i.e the terms containing $\eta^1$ still arise only as boundary terms evaluated at the end-points of the geodesic (section \ref{VOL}). So the only requirement is that $\eta^1$ vanishes at the end points of the geodesic.  

\section{Variation of geodesic length}\label{VOL}
The calculation of the variation of the geodesic is prototypical of the case where there are no metric perturbations. However to our knowledge  the extra terms arising due to metric perturbations have not been considered before. The action for a geodesic and the first variations is given by\footnote{G not to be confused with the gravitational constant.},
\beq
S&=&\int\underbrace{\sqrt{g_{\mu\nu}\frac{dx^\mu}{d\tau}\frac{dx^\nu}{d\tau}}}_Gd\tau\\
\delta S&=&\int\frac{1}{2\sqrt G}\left[2g_{\mu\nu}\frac{dx^\mu}{d\tau}\frac{d\delta x^\nu}{d\tau}+\delta g_{\mu\nu}\frac{dx^\mu}{d\tau}\frac{dx^\nu}{d\tau}\right]d\tau
\eeq
Note that $\delta g_{\mu\nu}(x)=\left(\frac{\partial g_{\mu\nu}(x)}{\partial x^\rho}\delta x^\rho+\tilde\delta g_{\mu\nu}(x)\right)$, where $\tilde\delta g_{\mu\nu}$ is the change in $g_{\mu\nu}$ purely due to metric perturbations. The following identifications will be made in the spirit of a Taylor expansion (appendix \ref{SOD}), $\delta x^\sigma=\accentset{(1)}\eta~^\sigma$,~$\delta^2 x^\sigma=\accentset{(2)}\eta~^\sigma$,~$\tilde\delta g_{\mu\nu}=\accentset{(1)}{h}_{\mu\nu}$,~$\tilde\delta^2 g_{\mu\nu}=\accentset{(2)}{h}_{\mu\nu}$. In the absence of metric perturbations on ends up with the equation for a curve to be geodesic.
\beq
\delta S&=&\int\frac{1}{2\sqrt G}\left[\frac{d}{d\tau}\left(2g_{\mu\nu}\frac{dx^\mu}{d\tau}\delta x^\nu\right)-2\frac{\partial g_{\mu\nu}}{\partial x^\rho}\frac{dx^\mu}{d\tau}\frac{dx^\rho}{d\tau}\delta x^\nu \right.\nn
&&\hspace{1cm}\left.-2g_{\mu\nu}\frac{d^2x^\mu}{d\tau^2}\delta x^\nu+\frac{\partial g_{\mu\nu}}{\partial x^\rho}\frac{dx^\mu}{d\tau}\frac{dx^\nu}{d\tau}\delta x^\rho+\tilde\delta g_{\mu\nu}\frac{dx^\mu}{d\tau}\frac{dx^\nu}{d\tau}\right]d\tau
\eeq
On the initial curve $G$ can be set to one assuming that the curve is affinely parametrized and therefore the geodesic equation follows if metric perturbation is zero i.e,
\beq
-2\frac{\partial g_{\mu\nu}}{\partial x^\rho}\frac{dx^\mu}{d\tau}\frac{dx^\rho}{d\tau}-2g_{\mu\nu}\frac{d^2x^\mu}{d\tau^2}+\frac{\partial g_{\mu\rho}}{\partial x^\nu}\frac{dx^\mu}{d\tau}\frac{dx^\rho}{d\tau}=G_\nu=0
\eeq
We move on and calculate the second variation of the action,
\beq
\delta^2S&=&\int\frac{1}{2\sqrt G}\left[\underbrace{2\delta g_{\mu\nu}\frac{dx^\mu}{d\tau}\frac{d\delta x^\nu}{d\tau}}_I+\underbrace{2g_{\mu\nu}\frac{d\delta x^\mu}{d\tau}\frac{d\delta x^\nu}{d\tau}}_{II}+\underbrace{2g_{\mu\nu}\frac{dx^\mu}{d\tau}\frac{d\delta^2 x^\nu}{d\tau}}_{III}+\underbrace{\delta^2 g_{\mu\nu}\frac{dx^\mu}{d\tau}\frac{dx^\nu}{d\tau}}_{IV}\right.\nn
&&\left.~~~~~~~~~~~~+\underbrace{2\delta g_{\mu\nu}\frac{dx^\mu}{d\tau}\frac{d\delta x^\nu}{d\tau}}_{V}\right]d\tau-\int\frac{1}{4 G^{3/2}}\underbrace{\left[2g_{\mu\nu}\frac{dx^\mu}{d\tau}\frac{d\delta x^\nu}{d\tau}+\delta g_{\mu\nu}\frac{dx^\mu}{d\tau}\frac{dx^\nu}{d\tau}\right]^2}_{VI}d\tau
\eeq
Since these expressions are evaluated on the initial curve, which is an affinely parametrized geodesic, we must set $G=1$. To evaluate term $IV$ we note the following,
\beq
\delta^2g_{\mu\nu}&=&\delta\left(\frac{\partial g_{\mu\nu}}{\partial x^\rho}\delta x^\rho+\tilde\delta g_{\mu\nu}\right)=\frac{\partial^2g_{\mu\nu}}{\partial x^\rho\partial x^\sigma}\delta x^\rho\delta x^\sigma+2\frac{\partial\tilde\delta g_{\mu\nu}}{\partial x^\rho}\delta x^{\rho}+\frac{\partial g_{\mu\nu}}{\partial x^\rho}\delta^2 x^\rho+\tilde\delta^2g_{\mu\nu}
\eeq
Note that the $\delta^2x^\rho$ term in $IV$ and term $III$ together can be written as,
\beq
\int\frac{1}{2\sqrt G}\left[\frac{d}{d\tau}\left(2g_{\mu\nu}\frac{dx^\mu}{d\tau}\delta^2 x^\nu\right)+G_\nu\delta^2 x^\nu\right]d\tau
\eeq
The total derivative term in the above can again be ignored and the second term is zero by the equation of geodesic.

Term $VI$ is just the square of the first variation and therefore can be written as
\beq
\int\frac{1}{4 G^{3/2}}\left[\frac{d}{d\tau}\left(2g_{\mu\nu}\frac{dx^\mu}{d\tau}\delta x^\nu\right)+G_\nu\delta x^\nu+\tilde\delta g_{\mu\nu}\frac{dx^\mu}{d\tau}\frac{dx^\nu}{d\tau}\right]^2d\tau
\eeq
The second term in the above is again zero by geodesic equation.

Terms $I$, $V$ and $II$ together can be re ordered and ignoring the total derivative terms, can be written as \footnote{~~~$\accentset{(1)}\eta~^1$ is the deviation vector in the direction of the tangent vector to the geodesic. For timelike geodesics however one must use $\accentset{(1)}\eta~^0$. The results however does not depend on this identification.}
\begin{gather}
\frac{1}{\sqrt G}\left[\accentset{(1)}C^{\alpha}_{\mu\nu}\frac{dx^\mu}{d\tau}\frac{dx^\nu}{d\tau}\accentset{(1)}{\eta_\alpha}-\partial_\sigma f(x)\eta^\sigma\accentset{(1)}\eta~^1\right],
\end{gather}
where \cref{IJE} has been used. The $\tilde\delta g_{\mu\nu}$ terms upto total derivatives give,
\begin{gather}
\frac{1}{\sqrt G}\left[-2\accentset{(1)}C^{\alpha}_{\mu\nu}\frac{dx^\mu}{d\tau}\frac{dx^\nu}{d\tau}\accentset{(1)}{\eta_\alpha}+\frac{1}{2}\accentset{(2)}{h}_{\mu\nu}\frac{dx^\mu}{d\tau}\frac{dx^\nu}{d\tau}\right]
\end{gather}
The final expression obtained is given by,
\begin{gather}\label{S1}
\delta S=\int\frac{1}{2}\left[
\accentset{(1)}{h}_{\mu\nu}\frac{dx^\mu}{d\tau}\frac{dx^\nu}{d\tau}\right]d\tau\end{gather}
\begin{align}\label{S2}
\delta^2 S&=
-\int\frac{1}{4}\left[\frac{d}{d\tau}\left(2g_{\mu\nu}\frac{dx^\mu}{d\tau}\accentset{(1)}{\eta_\mu}\right)+\accentset{(1)}{h}_{\mu\nu}\frac{dx^\mu}{d\tau}\frac{dx^\nu}{d\tau}\right]^2d\tau\nn
&+\int\left[\underbrace{-\accentset{(1)}C^{\alpha}_{\mu\nu}\frac{dx^\mu}{d\tau}\frac{dx^\nu}{d\tau}\accentset{(1)}{\eta_\alpha}}_{A}-\partial_\sigma f(x)\accentset{(1)}\eta~^\sigma\eta^1+\frac{1}{2}~\accentset{(2)}{h}_{\mu\nu}\frac{dx^\mu}{d\tau}\frac{dx^\nu}{d\tau}\right]d\tau
\end{align}
As a final step we see that the second order variation is independent of $\eta^1$. Note that the square term in the above expression can be expanded to get two $\accentset{(1)}\eta~^1$ dependent terms viz.,
\beq
-\int\left[\left(\frac{d\accentset{(1)}\eta~^1}{d\tau}\right)^2+\frac{d\accentset{(1)}\eta~^1}{d\tau}~\accentset{(1)} h_{\mu\nu}\frac{dx^\mu}{d\tau}\frac{dx^\nu}{d\tau}\right]
\eeq
The above expression can be integrated by parts. Leaving a total derivative it gives $-F_1\accentset{(1)}\eta~^1+\partial_\sigma f(x)\accentset{(1)}\eta~^\sigma\accentset{(1)}\eta~^1$ which essentially cancels the $F_1\accentset{(1)}\eta~^1$ coming from term $A$ and $-\partial_\sigma f(x)\accentset{(1)}\eta~^\sigma\accentset{(1)}\eta~^1$ term in \cref{S2}.
The final expression for $\delta^2 S$ is given by,
\begin{gather}\label{gg}
\delta^2 S=
-\int\frac{1}{4}\left[\accentset{(1)}{h}_{\mu\nu}\frac{dx^\mu}{d\tau}\frac{dx^\nu}{d\tau}\right]^2d\tau+\int\left[F^A\accentset{(1)}{\eta_A}+\frac{1}{2}~\accentset{(2)}{h}_{\mu\nu}\frac{dx^\mu}{d\tau}\frac{dx^\nu}{d\tau}\right]d\tau,
\end{gather}
where $A$ is summed over $0$ and $2$.
\section{Solutions of ``generalized deviation equation'' for rotating $BTZ$ like perturbations about $AdS_3$}\label{BTZgeo}
Consider the $AdS_3$ metric
\beq
ds^2=\frac{dz^2-dt^2+dx^2}{z^2}
\eeq
The equation for a spacelike geodesic of maximal length, parametrized by $\tau$ is then given by
\begin{align}\label{eqgeo0}
\frac{dx}{d\tau}={z^2\over{z^{(0)}_{*}}}
\end{align}
\begin{align}\label{eqgeo}
\frac{dz}{d\tau}=\pm z\sqrt{1-\left(\frac{z}{z^{(0)}_{*}}\right)^2},
\end{align}
where ${z^{(0)}_{*}}$ is the $AdS$ turning point. Besides the plus sign denotes the half going into the bulk and the minus sign denotes the half of the geodesic approaching the boundary.
From \cref{eqgeo} we can obtain the size of the subsystem in terms of the $AdS$ turning point
\begin{gather}\label{kk}
 l=2 {z^{(0)}_{*}} \int_{0}^{1} dk {k\over\sqrt{1-k^2}}=2{z^{(0)}_{*}}
\end{gather}

In order to calculate the integrals in the previous section we need both halves of the geodesic. Though both halves of the geodesic are identical, except for a change in sign of the velocity, the deviations may undergo non-trivial changes. To account for this we continue the solution of the ingoing half of \cref{eqgeo} to negative values of the affine parameter setting it equal to zero at the turning point. Hence the parameter~$\tau~\epsilon~~(-\infty,\infty)$ now covers the full geodesic. Therefore the full curve is now a map $\gamma:(-\infty,\infty)\rightarrow\mathcal M$. The solution  given by:
\beq
z(\tau)={z^{(0)}_{*}}\sech(\tau),
\eeq
where we have fixed the constant of integration in such a way that $\tau=0$ at the turning point. This solution can be substituted in \cref{eqgeo0} to get a solution $x(\tau)$. 

The components of $C^{\mu}~_{\nu\rho}$ can be obtained with the expressions for $h^{(1)}_{\mu\nu}$ given in Appendix \ref{FG}. Note that,
\[
\accentset{(1)}{h}_{\mu\nu}=
  \begin{bmatrix}
    \frac{(r_+^2 + r_-^2)}{2} & ~~0~~ &-r_+ r_- \\
    0&0&0\\
    -r_+ r_- & 0 & \frac{(r_+^2 + r_-^2)}{2}
  \end{bmatrix}
\]
Denoting $a=\frac{(r_+^2 + r_-^2)}{2}$ and $b=-r_+ r_-$.
\begin{gather}
\accentset{(1)}{C} ^{t }_{z ~t }=-z ~a,~\accentset{(1)}{C}^{x }_{z ~t }=z ~b,~\accentset{(1)}{C} ^{t }_{t ~z }=-z ~a,~\accentset{(1)}{C} ^{x}_{t ~z }=z ~b\nn
\accentset{(1)}{C} ^{t }_{x ~z }=-z ~b,~\accentset{(1)}{C}^{x }_{x ~z }=z ~a,~{C} ^{t }~_{z ~x }=-z ~b,~\accentset{(1)}{C} ^{x }~_{z ~x }=z ~a
\end{gather}
The tetrads that are parallely propagated along the geodesic are given by,
\beq
e^\mu_0=(z,0,0),~~e^\mu_1=\left(0,\pm z\sqrt{1-\left(z\over{z^{(0)}_{*}}\right)^2},{z^2\over{z^{(0)}_{*}}}\right),~~e^\mu_2=\left(0,{z^2\over{z^{(0)}_{*}}},\mp z\sqrt{1-\left(z\over{z^{(0)}_{*}}\right)^2}\right)\nn
\eeq
We therefore only need to solve the first two of the generalized geodesic equation. The first equation can be recast as \footnote{We have removed the (1) superscript in $\accentset{(1)}\eta$ in this section. All $\eta$'s in this section correspond to first order deviation vector. It is however necessary to make this distinction in \cref{SOD}.},
\beq
(\eta^0)^{\prime\prime} -\eta^0 - {2\over{z^{(0)}_{*}} } ~b ~z^2 ~z^{\prime}=0
\eeq 
A general solution of this equation is given by:
\beq
\eta^0=C_1e^{\tau}+C_2e^{-\tau}+\frac{2 {z^{(0)}_{*}}^2 b e^{-\tau} (-1 - 2 e^{2 \tau} + 2 e^{4 \tau} + e^{6 \tau})}{3(1+e^{2\tau})^2}
\eeq
To deal with the pathological nature of the coordinates at $z=0$, we will put the boundary conditions $\eta^0(p)=\eta^0(-p)=0$ for some cutoff $p$ and take $p\rightarrow \infty$ in the integrals. With $C_1,C_2$ fixed in terms of $p$ the final solution becomes,
\beq
\eta^0=b [-\sech(p)^2 \sinh(\tau) + \sech(\tau) \tanh(\tau)]{z^{(0)}_{*}}^2
\eeq 
The equation for $\eta^2$ is,
\beq
(\eta^{2})^{\prime\prime}-\eta^{2} -{2\over{z^{(0)}_{*}} }~ a~ z~z^{\prime 2} =0.
\eeq
Similarly as stated above the complete solution with proper boundary conditions is,
\beq
\eta^2=\frac{8 {z^{(0)}_{*}}^2 ~a~~e^{4 p+3 \tau} [-(3 + \cosh{4 p}) \cosh{2 \tau} + \cosh{2 p}~(3 + \cosh{4 \tau})]}{3 (1 + e^{2 p})4 (1 + e^{2 \tau})^3}
\eeq
To calculate the integrals in \cref{S1,gg} we need an expression for $\accentset{(2)}{h}_{\mu\nu}$ which is obtained from $F-G$ expansion in \cref{FG}.\\
\[
\frac{\accentset{(2)}h_{\mu\nu}}{2}=
  z^2\begin{bmatrix}
     \frac{-(r_-^2-r_+^2)^2}{16} & 0 &0\\
     0&0&0\\
    0 &0& \frac{(r_-^2-r_+^2)^2}{16}
  \end{bmatrix}
\]

and then taking the $p\rightarrow \infty$ gives,
\beq
\delta S&=&\frac{2 {z^{(0)}_{*}}^2 a}{3}=\frac{l^2(r_+^2+r_-^2)}{48G}\\
\delta^2 S&=&{z^{(0)}_{*}}^4\left(-\frac{1}{4}\left(\frac{32 a^2}{35}\right)+\left(\frac{8b^2}{35}-\frac{8a^2}{63}+\frac{4(b^2-a^2)}{15}\right)\right)=-\frac{4{z^{(0)}_{*}}^4(b^2+a^2)}{45}\nn
&=&-\frac{l^4}{720G}\left({(r_+^2 +r_-^2)^2 +4 r_+^2 r_-^2\over 4}\right)\nn
\eeq
Therefore the total change in entanglement entropy upto second order is,
\begin{align}\label{ok1}
\Delta S_E=&\frac{1}{4G}\left[\delta S+\frac{1}{2}\delta^2 S\right]\nn
=&\frac{l^2(r_+^2+r_-^2)}{48G}-\frac{l^4}{1440G}\left({(r_+^2 +r_-^2)^2 +4 r_+^2 r_-^2\over 4}\right).
\end{align}
In the next section we will verify this expression by deriving it from the expression of HEE for rotating BTZ obtained by HRT proposal.

Note: It is important to  note that when the perturbed metric is static there will be no off diagonal $h_{tx}$ like terms in the perturbation, 
and therefore the time component of the deviation vector will be trivial. For example in the non rotating BTZ case $r_-\rightarrow 0,r_+\rightarrow M$ i.e $b=0$. 
Hence for non `rotating BTZ' like perturabtions the time component of the deviation vector is zero. Hence the perturbed curve is still on a $t=constant$ slice.
Thus the information regarding different proposals (viz RT and HRT) is already incorporated in the deviation vector and the perturbed metric. 
Hence both can be addressed using this construction. 


\section{Entanglement First Law}\label{FL}
It has been shown in  \cite{Wong} that the change in  entanglement entropy $(\delta S_A)$ of a subregion $A$, under a small perturbations of the density matrix $\rho=\rho^0 + \delta\rho$ of a pure state in QFT,
satisfies a local first law of entanglement thermodynamics viz
\begin{align}
 \delta S_A (x)=\beta_{0}\delta E_{A}.
\end{align}
Where $\delta E_A =\delta<\hat{T}_{00}>Vol(A)$ is the excitation energy and $\beta_0 ={\int_A \beta(x)\over Vol(A)}$ is the average inverse entanglement temperature inside $A$. 
The density matrix for a mixed state at finite temperature $T$ and conserved charge $Q_a$ and chemical potential $\mu_a$ has the following form
\begin{gather*}
 \rho=\frac{\exp\left(-\frac{(H-\mu_a Q_a)}{T}\right)}{Z}
\end{gather*}
The first law gets modified to
\begin{gather}
 \delta S_A (x)=\beta_{0}\left(\delta E_{A}-\mu_a \delta Q_{aA} \right).
\end{gather}
Where $\delta Q_{aA}=\delta <Q_a>Vol(A)$.
 For rotating BTZ background corresponding density matrix is given by
\begin{align}
 \rho={exp-\beta\left(H-\Omega J\right)\over Z}.
\end{align}
Where $H$ and $J$ are the hamiltonian and angular momentum of the CFT and $\Omega$ is the angular velocity (which is essentially a chemical potential for the conserved angular momentum). 
According to the above treatment a simillar first law expression should also hold with $\mu_a$ replaced with $\Omega_A$ and $Q_A$ replaced with $J_A$.

Now we derive the entanglement first law for rotating BTZ geometry using the expression for HEE.
Holographic entanglement entropy for rotating BTZ geometry is given by\cite{Hubeny}.\footnote{ A separate calculation for HEE  in terms of geodesic length can be found in Appendix \ref{BTZ}}
\begin{gather}
 S_E={c\over 6}\ln\Biggl({\beta_+ \beta_-\over \pi^2 \epsilon^2} \sinh\left({\pi l \over \beta_+}\right)\sinh\left({\pi l \over \beta_-}\right)\Biggr)
\end{gather}
where $\beta_{\pm}={2\pi\over r_+ \pm r_-}$ are the inverse temperature for left and right moving modes  and $l$ is the size of the subsystem in the dual $CFT_2$. $\epsilon$ is the
UV cutoff and $c={3\over 2G}$ is the central charge of the dual $CFT_2$ and $G$ is the 3 dimensional Newton's constant. The increase in HEE of a subsystem of size $l$ is obtained by subtracting it from pure $AdS_3$ contribution given by
\begin{gather}
 S_{AdS_3}={c\over 3}\ln\Biggl({l\over\epsilon}\Biggr).
\end{gather}
For rotating BTZ geometry  the increase in HEE of a subsystem of size $l$ is given by
\begin{gather}\label{p4.3}
 \Delta S_E=S_E - S_{AdS_3}\\\notag
 ={c\over 6}\ln\Biggl({\beta_+ \beta_-\over \pi^2 l^2} \sinh\left({\pi l \over \beta_+}\right)\sinh\left({\pi l \over \beta_-}\right)\Biggr).
\end{gather}


The physical thermodynamic observables of the dual $CFT_2$ can be obtained by expanding the rotating BTZ geometry in 
suitable Fefferman-Graham (asymptotic) coordinates near the $AdS$ boundary \cite{fg}, given in the appendix \ref{FG}. These are summarized here. 
The energy and angular momentum for the strip subsystem are

\begin{align}\label{EA}
 \Delta E={1\over 8\pi G} l {(r_+^2 +r_-^2)\over 2}={\pi l\over 8 G}\left({1\over\beta_+^2}+{1\over\beta_-^2}\right)\nn
 \Delta J={1\over 8\pi G} l (r_+ r_-)={\pi l\over 8 G}\left({1\over\beta_-^2}-{1\over\beta_+^2}\right).
\end{align}
The entanglement temperature has been defined in and \cite{JT}. By the same argument one should be able to define an entanglement angular velocity. 
\begin{gather}
 {1\over T_E}={\partial(\Delta S_E)\over \partial(\Delta E)}\bigg |_ {l=fixed},~~~~~{\Omega_E\over T_{E}}={\partial(\Delta S_E)\over \partial(\Delta J)}\bigg |_{l=fixed}
\end{gather}
Using eq. \ref{p4.3}, \ref{EA} we get
\begin{gather}\label{1}
 {1\over T_E}=-{{(\beta_+^2 +\beta_-^2)-l\pi(\beta_+\coth{\pi l\over\beta_+}+\beta_-\coth{\pi l\over\beta_-})}\over 2 l\pi}\\
 {\Omega_E\over T_E}={{\beta_+^2 -\beta_-^2 -l\pi(\beta_+\coth{\pi l\over\beta_+}-l\pi\beta_-\coth{\pi l\over\beta_-})}\over 2 l\pi}
\end{gather}
Using these definition it is quite logical to write an alternative form of the first involving differential changes in $\Delta S$. This first law is valid upto all orders in subsystem size.
\begin{gather}\label{2}
 d(\Delta S_E)={1\over T_E}d(\Delta E)-{\Omega_E\over T_E}d(\Delta J)
\end{gather}
In the above form $\Delta S$ must be interpreted as subtracting a ground state entropy (pure AdS)from the entropy of the excited state. The diffrential changes $d\Delta S$ are changes due to changes of the excited state itself. Hence the first law relates the change in $\Delta S$ due to changes in the BTZ parameters.

In the case of black hole thermodynamics the change in the entropy of a black hole is related to changes in the black hole parameters as one moves from one black hole solution to another in the phase space of solutions. Hence the above first is closer in spirit to the first law for black hole thermodynamics. In fact we further show that it is this first law that asymptotes to the first law for the BTZ black hole in the large system size limit.
As the derivatives don't act on $l$, \cref{2} can be written in terms of mass $(M={(r_+^2 + r_-^2)\over 8 G})$ and angular momentum  $(J={r_+ r_-\over 4 G})$ of rotating BTZ black hole as follows
\begin{gather}\label{3}
  d({2 G\pi\Delta S_E\over l})={1\over T_E}d(M)-{\Omega_E\over T_E}d(J)
\end{gather}
Taking the large subsystem size $l$ limit these quantites approaches their respectives thermal values
\begin{gather}
 \lim_{l\longrightarrow\infty} {2\pi\Delta S_E\over l}={\pi^2\over 2 G}\left({1\over\beta_+}+{1\over\beta_-}\right)={\pi r_+\over 2G}\\
 \lim_{l\longrightarrow\infty} T_E={2\over (\beta_+ +\beta_-)}={r_+^2-r_-^2\over 2\pi r_+}\\
  \lim_{l\longrightarrow\infty} \Omega_E={(\beta_+ -\beta_-)\over(\beta_+ +\beta_-)}={r_-\over r_+}
\end{gather}.
Thus \cref{3} approaches the first law of black hole thermodynamics in the large subsystem size ($l$) limit. 

In the small subsystem size limit ${\pi l\over \beta_\pm}\ll 1$,  
we can expand entanglement temperature and angular velocity upto second order in ${\pi l\over \beta_\pm}$ using \cref{1} 
\begin{gather}\label{bb}
 {1\over T_E}={\pi l\over 3}-{\pi l^3\over 90}\left({1\over\beta_+^2}+{1\over\beta_-^2}\right)+\cdots\\
 \Omega_E={\pi^2 l^2\over 15}\left({1\over\beta_-^2}-{1\over\beta_+^2}\right)+\cdots.
\end{gather}
Thus at leading order the entanglement temperature is inversely proportional to the subsystem size. The entanglement angular velocity at leading order is proportional to $({\pi l\over \beta_\pm})^2$ these 
contributions to the change in HEE appears only at second order and are due to second order gravitational perturbation and first order perturbations of the extremal surface. Dependence of 
entanglement temperature and angular velocity on subsystem size is given in fig(2) and fig(3). Note that perturbation of the entanglement temperature in CFT has been disscussed in \cite{Levine:2015uka} for example.
\vspace{1cm}

\begin{figure}[h]
  \begin{minipage}[h]{0.4\textwidth}
    
    \begin{overpic}[width=\linewidth]{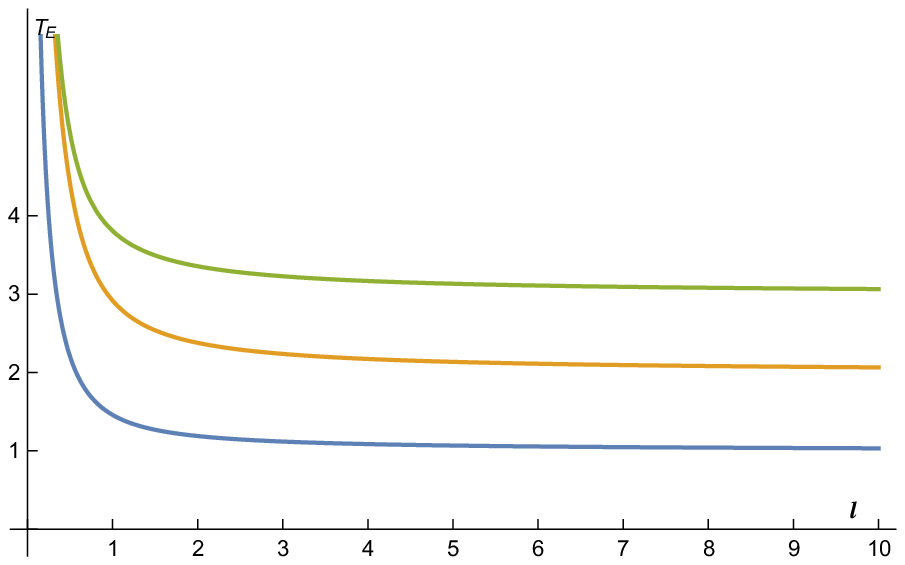}
\put(0,65){\large$T_{E}$}
\put(80,33){\small$T_{th}=3$}
\put(80,23){\small$T_{th}=2$}
\put(80,14){\small$T_{th}=1$}
\put(105,0){\Large$l$}
\end{overpic}
\caption{\it Plot of $T_E$ v.s $'l'$ for different black hole temperatures. As $'l'$ increases the temperature asymptotes to thermal value}
  \end{minipage}
  \hfill
  \begin{minipage}[h]{0.4\textwidth}
    \begin{overpic}[width=\linewidth]{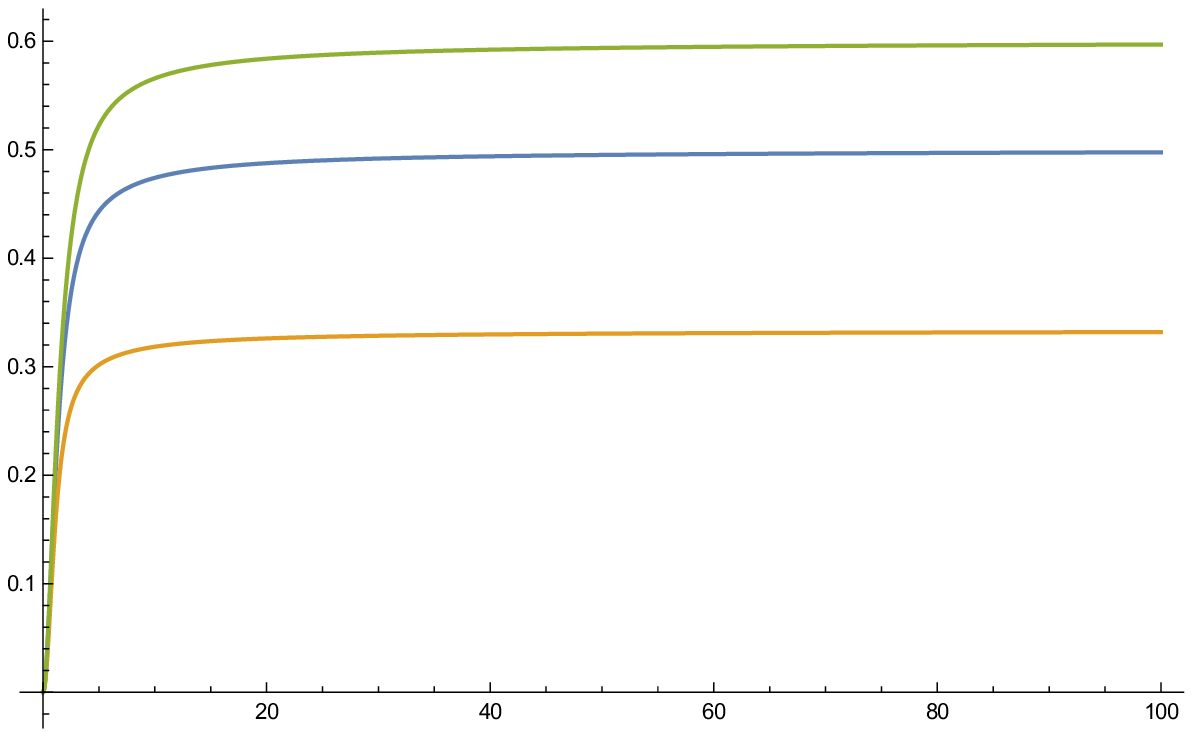}
\put(0,65){\large$\Omega_{E}$}
\put(80,35){\small$\Omega_{th}={1\over 3}$}
\put(80,50){\small$\Omega_{th}={1\over 2}$}
\put(80,60){\small$\Omega_{th}={3\over 5}$}
\put(105,0){\Large$l$}
\end{overpic}
    \caption{\it Plot of $\Omega_E$ v.s $'l'$ for different blackhole angular velocity. As $'l'$ increases the angular velocity asymptotes to thermal value}
  \end{minipage}
\end{figure}

Similarly we can expand \cref{p4.3} in the small subsystem size limit ${\pi l\over \beta_\pm}\ll 1$
\begin{gather}\label{cc}
\Delta S_E=\Delta S_1+\Delta S_2=\frac{l^2(r_+^2+r_-^2)}{48G}-\frac{l^4}{1440G}\left({(r_+^2 +r_-^2)^2 +4 r_+^2 r_-^2\over 4}\right)
\end{gather}
It is important to note that this expression exactly matches with equation \cref{ok1} reproduced earlier by studying geodesic deviations.It is important to note that although full expression for $\Delta S$ is known for rotating BTZ geometry. This is not the case for other backgrounds in higher dimensions. In those cases expression for $\Delta S$ is obtained perturbatively. In \cref{BTZgeo} we gave a prescription in $2+1$ to calculate $\Delta S$ by accounting for first order changes in the minimal surface and second order gravitational perturbations. Here we verify our result for rotating BTZ case.
 Further in the large $l$ limit the ratio $2\pi\Delta S\over l$ approaches
the Bekenstein Hawking entropy as shown in fig(4).
\begin{figure}[h]
 \centering
 \begin{overpic}[width=8cm]{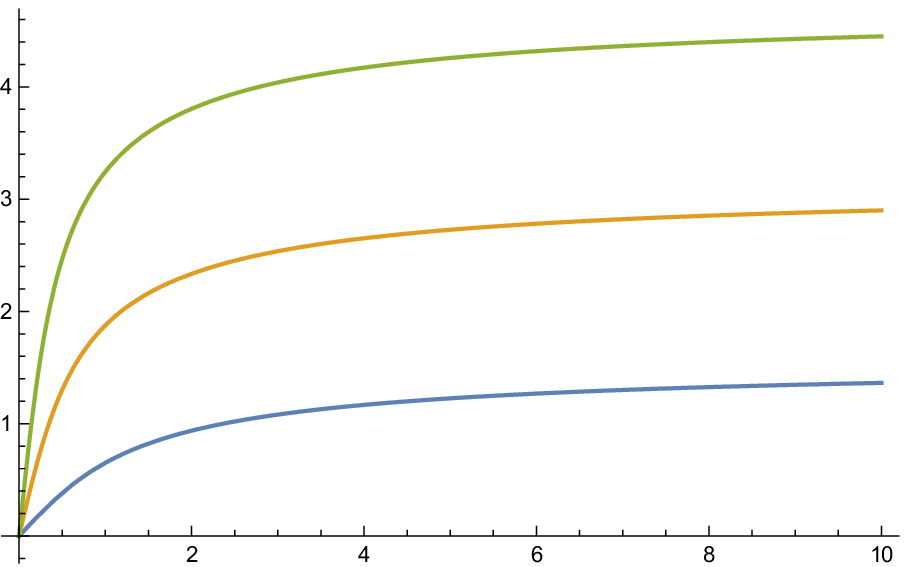}
\put(5,60){\large$\frac{2G\pi\Delta S_E}{l}$}
\put(80,22){\small$S_{th}=\pi^2$}
\put(80,41){\small$S_{th}=2\pi^2$}
\put(80,60){\small$S_{th}=3\pi^2$}
\put(105,0){\Large$l$}
\end{overpic}
 \caption{\it Plot of $\frac{2G\pi\Delta S_E}{l}~~\text{vs}~~ l$ for different Bekenstein Hawking entropy. As $'l'$ increases $\frac{2G\pi\Delta S_E}{l}$ asymptotes to thermal value }
\end{figure}

Now at first order from \cref{bb,cc} we can write the first law as
\begin{gather}\label{dd}
 d(\Delta S_1)={1\over T^{(1)}_E}d(\Delta E).
\end{gather}
Where  $T^{(1)}_E={3\over\pi l}$ is the entanglement temperature at first order. Thus at first order \cref{dd} can be integrated to give
\begin{gather}
 \Delta S_1={1\over T^{(1)}_E}\Delta E.
\end{gather}
Which is the entanglement first law obtained in \cite{JT,Caputa}. However at second order \cref{2} can not be integrated as $T_E$ also depends on details of excitation. Hence at second order one can atmost write an inequality
\begin{gather}\label{mm}
 \Delta S_E<{1\over T_E}\Delta E-{\Omega_E\over T_E}\Delta J.
\end{gather}
 It will be interesting to check whether if  expression \cref{mm} resembles in spirit to the Bekenstein bound for rotating bodies \cite{Bekenstein:1980jp,Hod:2000ju, Bekenstein:1999cf}, or the Penrose inequality for axis symmetric spacetimes \cite{Dain:2011mv,Park:2015hcz}. The QFT analogue of the Bekenstein bound for non rotating bodies was holographically verified in \cite{Blanco}.
\section{Discussions}\label{CON}
We conclude that  second order changes in $\Delta S_E$ for $2+1$ dimensions correspond to \begin{enumerate} 
\item Second order gravitational perturbations and 
\item First order changes in the shape of extremal surface. 
\end{enumerate}
The second order gravitational perturbations can be obtained by solving the perturbed Einstein's equation. Alternatively when the bulk metric is known, this corresponds to the $\mathcal O(z^4)$ (in $2+1$ dimensions) terms in the Fefferman Graham expansion. However a systematic approach for finding the change in shape of the extremal surface is not known. We propose that these changes can be systematically calculated by solving the "generalized deviation equation". We futher write an alternative form of the first law for entanglemnt thermodynamics given by,
\begin{gather}
 d(\Delta S_E)={1\over T_E}d(\Delta E)-{\Omega_E\over T_E}d(\Delta J)
\end{gather}
This has been shown to asymptote exactly to the black hole first law for BTZ in the large system size limit.

Having obtained a covariant expreesion for the change in entanglement entropy upto second order in the perturbation series, it will be interesting to check what constraints Einstein's equation (second order linearized Einstein's equation to be precise) puts on the dynamics of $\Delta S$ as done in \cite{Nozaki,Jyotirmoy,Lin:2014hva}. Moreover one may
attempt to follow the procedure outlined in this paper for time dependent
perturbations over $AdS_{3}$, the CFT calculation of which has been per-
formed in \cite{Sheikh-Jabbari:2016znt}.

Further generalization to higher dimensions seems to be a plausible work that we would like to address in the future. This would provide one with a definite presription for calculation of change in entanglement entropy perturbatively in higher dimension. 
\section*{Acknowledgements}
We would like to thank Prof. Amit Ghosh and Prof. Harvendra Singh for useful discussions. RM is financially supported by DAE Government of India. AG acknowledges the hospitality of Theory Division, Saha Institute of Nuclear Physics where the work was done. 

\newpage
\appendix
\section{Fefferman Graham Expansion, boundary stress tensor and perturbations about $AdS_3$}\label{FG}
The rotating BTZ metric is given as,
\begin{gather}\label{mBTZ}
ds^2=-\frac{(r^2-r_+^2)(r^2-r_-^2)}{r^2}~dt^2
+\frac{r^2}{(r^2-r_+^2)(r^2-r_-^2)}~dr^2+r^2\left(d\phi-\frac{r_+r_-}{r^2}~dt\right)^2,
\end{gather}
where $r_+,r_-$ are the radii of the outer and inner horizon respectively. The physical observables like energy and angular momentum can be obtained by expanding the above metric in suitable Fefferman Graham(Asymptotic) coordinates near the $AdS_3$ boundary. 
This can be realized by defining a new coordinate $\rho$ through $\frac{d\rho}{\rho}=\frac{dr}{r\sqrt{f(r)}}$. In terms of $\rho$ this metric becomes

\begin{gather}
 ds^2=\frac{d\rho^2}{\rho^2}+\rho^2 \Biggl[(-dt^2+d\phi^2)+\frac{1}{\rho^2}\Biggl(\frac{(r_+^2 + r_-^2)}{2}dt^2 - 2 r_+ r_- dt d\phi+\frac{(r_+^2 + r_-^2)}{2}d\phi^2\Biggr)\nn
 +\frac{1}{\rho^4}\Biggl(\frac{-(r_-^2-r_+^2)^2}{16}dt^2+\frac{(r_-^2-r_+^2)^2}{16}d\phi^2\Biggr)\Biggr]
\end{gather}
In coordinates $(\rho=\frac{1}{z})$ the metric becomes
\begin{gather}
 ds^2={dz^2+(\eta_{\mu\nu} +z^2 \gamma^{(2)}_{\mu\nu} +z^4 \gamma^{(4)}_{\mu\nu}+...)dx^\mu dx^\nu\over z^2}
\end{gather}
The above metric is now in Fefferman Graham form.  

Where the boundary energy momentum tensor ($\langle T_{\mu\nu}\rangle={d\over 16\pi G}\gamma^{(d)}_{\mu\nu}$ in $d+1$ dimensions) \cite{Skenderis:1999nb,deHaro:2000vlm} is given by
\[
8\pi G~T_{\mu\nu}=\gamma^{(2)}_{\mu\nu}=
  \begin{bmatrix}
    \frac{(r_+^2 + r_-^2)}{2} & -r_+ r_- \\
    -r_+ r_- & \frac{(r_+^2 + r_-^2)}{2}
  \end{bmatrix}
\]

In $2+1$ dimensions there are no conformal anomalies in the stress tensor  \cite{Skenderis:1999nb} and $\gamma^{(4)}_{\mu\nu}$ is given by
\[
\gamma^{(4)}_{\mu\nu}=
  \begin{bmatrix}
     \frac{-(r_-^2-r_+^2)^2}{16} & 0 \\
    0 &\frac{(r_-^2-r_+^2)^2}{16}
  \end{bmatrix}\]

\section{Length of space-like geodesic for rotating BTZ} \label{BTZ}
According to HRT proposal extremal surfaces in 2+1 dimensions are given by spacelike  geodesics. Here we obtain HEE for rotating BTZ black hole by calculating geodesic length without using the fact that BTZ is locally isometric to $AdS_3$. In the rotating BTZ metric \cref{mBTZ}, we will introduce the 
following notations \footnote{ These notations should be read independently of those introduced in the body of the paper. However the final expression is obtained in terms of quantities which have been introduced earlier.}\cite{Cruz:1994ir},
\begin{gather}
\mathscr M=r_+^2+r_-^2,~\mathscr J=2r_+r_-,~\beta_{\pm}=\frac{2\pi}{r_+\pm r_-},
\end{gather}
For a general curve parametrized by $\lambda$ (say) such that the tangent vector $(v^a)$ is given by $v^a=\left(\frac{dt}{d\lambda},\frac{dr}{d\lambda},\frac{d\phi}{d\lambda}\right)$, one has,
\begin{gather}
-m^2=-\frac{(r^2-r_+^2)(r^2-r_-^2)}{r^2}~\dot t^2+\frac{r^2}{(r^2-r_+^2)(r^2-r_-^2)}\dot r^2
+r^2\left(\dot \phi-\frac{r_+r_-}{r^2}~\dot t\right)^2,
\end{gather}
where dots imply derivative with respect to $\lambda$. If the curve is a geodesic and $\partial_t$ and $\partial_\phi$ being Killing vectors, one has the following constants of motion.
\beq
\mathscr E=-g_{ab}v^a\xi^b=\left(-\mathscr M+r^2\right)\dot t+\frac{\mathscr J}{2}\dot\phi,
\eeq
where $\xi^a=\partial_t^a$, and 
\beq
\mathscr L=g_{ab}v^a\Phi^b=r^2\dot\phi-\frac{\mathscr J}{2}\dot t,
\eeq
where $\Phi^a=\partial_\phi^a$. We define the following dimensionless coordinates and parameters for brevity.
\beq
\hat r=\frac{r}{\sqrt {\mathscr M}},~~~~\hat\phi=\phi\sqrt {\mathscr M},~~~~\hat t=t\sqrt {\mathscr M}\nn
\hat {\mathscr E}=\frac{\mathscr E}{\sqrt{\mathscr M}},~~~~\hat {\mathscr L}=\frac{\mathscr L}{\sqrt {\mathscr M}},~~~~\hat {\mathscr J}=\frac{\mathscr J}{\mathscr M}
\eeq
If the parameter $\lambda$ is taken to be the length along the geodesic, then following equations of motion follow for space-like geodesics ($m^2=-1$).
\beq
r^2\dot r^2&=&\left(r^4-r^2+\frac{\mathscr J^2}{4}\right)+\left(\mathscr E^2-\mathscr L^2\right)r^2+\mathscr L^2-\mathscr J\mathscr E \mathscr L\nn
\dot\phi&=&\frac{(r^2-1)L-\frac{1}{2}\mathscr J\mathscr E}{(r^2-r_+^2)(r^2-r_-^2)}\nn
\dot t&=&\frac{\mathscr Er^2-\frac{1}{2}\mathscr J\mathscr L}{(r^2-r_+^2)(r^2-r_-^2)}
\eeq
where we have omitted the hat from the quantities. 

It is easy to note that in the limit $r\rightarrow\infty$,
\beq
\frac{dt}{d\phi}\approx\frac{\mathscr E}{\mathscr L}
\eeq
The geodesic will penetrate most into the bulk if this is zero. This precisely implies that $\mathcal E=0$ \cite{Hubeny:2012ry}. Therefore with the substitution $u=r^2$ the radial equation reduces to

\beq
\frac{1}{4}\left(\frac{du}{d\lambda}\right)^2&=&u^2-(1+\mathscr L^2)u+\left(\mathscr L^2+\frac{\mathscr J^2}{4}\right)\nn
&=&(u-a)(u-b),
\eeq
where the following relations hold for $a$ and $b$.
\beq
a+b=(1+\mathscr L^2),~ab=\mathscr L^2+\frac{\mathscr J^2}{4}
\eeq
Without loss of generality, we can take $a$ to be the greater of the two roots and therefore the turning point of the geodesic. Then the geodesic length can be obtained as,
\beq
\Lambda&=&\int_a^{u_\infty}\frac{du}{\sqrt{(u-a)(u-b)}}\nn
&=&\log\left(\frac{4r^2_\infty}{a-b}\right)
\eeq

To express the above in terms of the subsystem size one has to relate $a-b$ to the subsystem size. We note that,
\beq
\frac{d\phi}{du}&=&\frac{1}{2}\frac{(u-1)\mathscr L}{(u-u_+)(u-u_-)\sqrt{(u-a)(u-b)}}
=\frac{1}{2}\left[\frac{A}{u-u_+}+\frac{B}{u-u_-}\right]\frac{1}{\sqrt{(u-a)(u-b)}},
\eeq
where $A=-\frac{u_-\mathscr L}{u_+-u_-},~B=\frac{u_+\mathscr L}{u_+-u_-}$~\footnote{Note that in these coordinates $u_++u_-=1$ and $u_+u_-=\frac{\mathscr J^2}{4}$}.

Now each of the integrals are of the form,
\begin{gather}
\int\frac{dx}{(x-c)\sqrt{(x-a)(x-b)}}\nn=\frac{1}{\sqrt{(a-c)(b-c)}}
\times\log\left[\frac{\sqrt{(a-c)(b-c)}(c-x)}{a(-2b+c+x)-2\sqrt{(a-c)(b-c)(x-a)(x-b)}-2cx+b(c+x)}\right]\nn
\end{gather}
Putting the limits,
\beq
\int_a^\infty\frac{dx}{(x-c)\sqrt{(x-a)(x-b)}}=\frac{1}{\sqrt{(a-c)(b-c)}}\log\left[\frac{\sqrt{(a-c)}+\sqrt{(b-c)}}{\sqrt{(a-c)}-\sqrt{(b-c)}}\right]
\eeq
Therefore the $\dot\phi$ integral gives, 
\begin{multline}\label{phidot}
l=\frac{A}{u_+-u_-}\log\left(\frac{\sqrt{(a-u_+)}+\sqrt{(b-u_+)}}{\sqrt{(a-u_+)}-\sqrt{(b-u_+)}}\right)
+\frac{B}{u_+-u_-}\log\left(\frac{\sqrt{(a-u_-)}+\sqrt{(b-u_-)}}{\sqrt{(a-u_-)}-\sqrt{(b-u_-)}}\right),
\end{multline}

where $l$ is the subsystem size. We need to impose a further restriction. Note that,
\beq
\frac{dt}{du}&=&-\frac{1}{4}\frac{\mathscr J\mathscr L}{(u-u_+)(u-u_-)\sqrt{(u-a)(u-b)}}
=\frac{1}{4}\left[\frac{C}{u-u_+}+\frac{D}{u-u_-}\right]\frac{1}{\sqrt{(u-a)(u-b)}},
\eeq
where $C=\frac{1}{u_+-u_-}$ and $D=-\frac{1}{u_+-u_-}$. Therefore the interval of time elapsed is given by,
\begin{multline}
\Delta T=\frac{\mathscr J}{2\sqrt {u_-}(u_+-u_-)}\log\left(\frac{\sqrt{(a-u_+)}+\sqrt{(b-u_+)}}{\sqrt{(a-u_+)}-\sqrt{(b-u_+)}}\right)\\
-\frac{\mathscr J}{2\sqrt {u_+}(u_+-u_-)}\log\left(\frac{\sqrt{(a-u_-)}+\sqrt{(b-u_-)}}{\sqrt{(a-u_-)}-\sqrt{(b-u_-)}}\right)
\end{multline}
Since the subsystem is on a constant $t$ slice on the boundary, the total elapsed time must be zero \cite{Hubeny:2012ry}. Therefore,
\beq
(\beta_++\beta_-)\tanh^{-1}\sqrt{{\frac{b-u_+}{a-u_+}}}=(\beta_+-\beta_-)\tanh^{-1}\sqrt{{\frac{b-u_-}{a-u_-}}}
\eeq
which can be re-ordered to give,
\begin{multline}\label{tdot}
\beta_+\left[\tanh^{-1}\sqrt{{\frac{b-u_-}{a-u_-}}}-\tanh^{-1}\sqrt{{\frac{b-u_+}{a-u_+}}}\right]=
\beta_-\left[\tanh^{-1}\sqrt{{\frac{b-u_-}{a-u_-}}}-\tanh^{-1}\sqrt{{\frac{b-u_+}{a-u_+}}}\right]
\end{multline}
Therefore using \cref{tdot} one has the following conditions on the solutions \cref{phidot},
\beq
\tanh\frac{\pi l}{\beta_+}=\frac{\sqrt{(b-u_-)(a-u_+)}-\sqrt{(b-u_+)(a-u_-)}}{\sqrt{(a-u_+)(a-u_-)}-\sqrt{(b-u_+)(b-u_-)}}\nn
\tanh\frac{\pi l}{\beta_-}=\frac{\sqrt{(b-u_-)(a-u_+)}+\sqrt{(b-u_+)(a-u_-)}}{\sqrt{(a-u_+)(a-u_-)}+\sqrt{(b-u_+)(b-u_-)}}
\eeq
From these one can get the following expression for $\sinh{\frac{\pi l}{\beta_+}}$ and $\sinh{\frac{\pi l}{\beta_-}}$.
\beq
\sinh^2\frac{\pi l}{\beta_+}=\frac{\left(\sqrt{(b-u_-)(a-u_+)}-\sqrt{(b-u_+)(a-u_-)}\right)^2}{(a-b)^2}\nn
\sinh^2\frac{\pi l}{\beta_-}=\frac{\left(\sqrt{(b-u_-)(a-u_+)}+\sqrt{(b-u_+)(a-u_-)}\right)^2}{(a-b)^2}
\eeq
Therefore,
\beq
\sinh^2\frac{\pi l}{\beta_+}\sinh^2\frac{\pi l}{\beta_-}=\frac{(u_+-u_-)^2}{(a-b)^2}
\eeq
Taking the positive square root we get,
\beq
\sinh\frac{\pi l}{\beta_+}\sinh\frac{\pi l}{\beta_-}=\frac{4\pi^2}{\beta_+\beta_-(a-b)}
\eeq
Therefore one can has the desired result for the geodesic length in terms of the subsystem size,
\beq
\Lambda=\log\left(\frac{\beta_+\beta_-}{\pi^2\epsilon^2}\sinh\left(\frac{\pi l}{\beta_+}\right)\sinh\left(\frac{\pi l}{\beta_-}\right)\right)
\eeq
where we have put $r_\infty=\frac{1}{\epsilon}$

To find the turning point in terms of the subsystem size, note that
\beq
\tanh\frac{\pi l}{\beta_+}\tanh\frac{\pi l}{\beta_-}=\frac{4\pi^2}{\beta_+\beta_-(a+b-1)}
\eeq In the small subsystem size approximation,
\begin{gather}
a+b\approx\frac{4}{l^2}\nn
a-b\approx\frac{4}{l^2}\nn
a\approx\frac{4}{l^2}
\end{gather}
So the turning point in the $r$ coordinate upto leading order is given by  
\begin{gather}\label{zz}
r_{*}\approx\frac{2}{l}
\end{gather}
\section{Second order deviations}\label{SOD}
Though we won't be using the solutions of the equation for the second order deviations, for completeness we do derive these expressions. Since we have seen that deviations along the geodesic does not effect the length it will suffice to take the perturbed geodesic to be affinely parametrized by the affine parameter of the unperturbed geodesic.
So consider the second order deviation from the $AdS_3$ geodesic viz. \footnote{The $\frac{1}{2}$ factor in $\accentset{(2)}{\eta}^{\mu}$ is introduced to interprete the expansion as a Taylor expansion in terms of a paramenter which is already included in the terms of the expansion.}
\beq\tilde{x^\mu}=x^{\mu}+\accentset{(1)}{\eta}^{\mu}+\frac{1}{2}\accentset{(2)}{\eta}^{\mu}\eeq
We write the second order perturbation of the metric as $\accentset{0}{g}_{ab}+\accentset{(1)}{h}_{ab}+\frac{1}{2}\accentset{(2)}{h}_{ab}$. The second order perturabation of the connection due to ``metric perturbations only" is given by,
\begin{gather}
\Tilde\Gamma^{\mu}_{\nu\rho}(x)=\accentset{0}{\Gamma}^{\mu}_{\nu\rho}(x)+\accentset{(1)}{C}^{\mu}_{\nu \rho}(x)-\accentset{(1)}{h}^{\mu\sigma}\left(\partial_\nu \accentset{(1)}{h}_{\rho\sigma}+\partial_\rho \accentset{(1)}{h}_{\nu \sigma}-\partial_\sigma \accentset{(1)}{h}_{\nu \rho}\right)
\nn+\frac{1}{2}\accentset{0}{g}^{\mu\sigma}\left(\partial_\nu \accentset{(2)}{h}_{\rho\sigma}+\partial_\rho \accentset{(2)}{h}_{\nu \sigma}-\partial_\sigma \accentset{(2)}{h}_{\nu \rho}\right)-\frac{1}{2}\accentset{(2)}{h}^{\mu \sigma}\left(\partial_\nu\accentset{0}{g}_{\rho \sigma}+\partial_\rho\accentset{0}{g}_{\nu \sigma}-\partial_\sigma\accentset{0}{g}_{\nu \rho}\right)\nn
=\accentset{0}{\Gamma}^{\mu}_{\nu\rho}(x)+\accentset{(1)}{C}^{\mu}_{\nu \rho}(x)+\accentset{(2)}{C}^{\mu}_{\nu \rho}(x)
\end{gather}
Therefore,
\begin{gather}
\tilde\Gamma^{\mu}_{\nu\rho}(\tilde{x})=\accentset{0}{\Gamma}^{\mu}_{\nu\rho}(x)+\partial_{\sigma}\accentset{0}{\Gamma}^{\mu}_{\nu\rho}(\accentset{(1)}\eta^{\sigma}+\accentset{(2)}\eta^{\sigma})+\accentset{0}{\Gamma}^{\mu}_{\nu\rho}\frac{d\accentset{(1)}\eta^{\nu}}{d\tau}+\accentset{(1)}C^{\mu}_{\nu\rho}(x)+\partial_{\sigma}\accentset{(1)}C^{\mu}_{\nu\rho}\accentset{(1)}\eta^{\sigma}\nn+\accentset{(2)}C^{\mu}_{\nu\rho}(x)+higher-order
\end{gather}
Putting this in the geodesic equation and collecting the 2nd order terms give,
\begin{gather}
\frac{d\accentset{(2)}\eta^{\mu}}{d\tau^2}+\accentset{0}{\Gamma}^{\mu}_{\nu\rho}\frac{d\accentset{(2)}\eta^{\nu}}{d\tau}\frac{dx^{\rho}}{d\tau}+\accentset{0}{\Gamma}^{\mu}_{\nu\rho}\frac{d\accentset{(2)}\eta^{\rho}}{d\tau}\frac{dx^{\nu}}{d\tau}+\partial_{\sigma}\accentset{0}{\Gamma}^{\mu}_{\nu\rho}\accentset{(2)}\eta^{\sigma}\frac{dx^{\nu}}{d\tau}\frac{dx^{\rho}}{d\tau}\nn
+\partial_{\sigma}\accentset{0}{\Gamma}^{\mu}_{\nu\rho}\accentset{(1)}\eta^{\sigma}\frac{dx^{\nu}}{d\tau}\frac{d\eta^{(1)\rho}}{d\tau}+\partial_{\sigma}\accentset{0}{\Gamma}^{\mu}_{\nu\rho}\accentset{(1)}\eta^{\sigma}\frac{d\accentset{(1)}\eta^{\nu}}{d\tau}\frac{dx^{\rho}}{d\tau}
+\accentset{(1)}{C}^{\mu}_{\nu\rho}\frac{dx^{\nu}}{d\tau}\frac{d\accentset{(1)}{\eta}^{\rho}}{d\tau}+\accentset{(1)}{C}^{\mu}_{\nu\rho}\frac{d\accentset{(1)}{\eta}^{\nu}}{d\tau}\frac{dx^{\rho}}{d\tau}\nn
+\partial_{\sigma}\accentset{(1)}C^{\mu}_{\nu\rho}\accentset{(1)}{\eta}^{\sigma}\frac{dx^{\nu}}{d\tau}\frac{dx^{\rho}}{d\tau}+\accentset{(2)}{C}^{\mu}_{\nu\rho}\frac{dx^{\nu}}{d\tau}\frac{dx^{\rho}}{d\tau}=0
\end{gather}
The first four terms can be identified as the left hand side of the Jacobi equation. Therefore,
\begin{gather}
\frac{\mathcal D^2\accentset{(2)}\eta~^{\mu}}{d\tau^2}+\mathcal R^{\mu}~_{\nu\rho\sigma}\frac{dx^\nu}{d\tau}\frac{dx^\sigma}{d\tau}\accentset{(2)}\eta~^{\rho}=-\left(\partial_{\sigma}\accentset{0}{\Gamma}^{\mu}_{\nu\rho}\accentset{(1)}\eta^{\sigma}\frac{dx^{\nu}}{d\tau}\frac{d\eta^{(1)\rho}}{d\tau}+\partial_{\sigma}\accentset{0}{\Gamma}^{\mu}_{\nu\rho}\accentset{(1)}\eta^{\sigma}\frac{d\accentset{(1)}\eta^{\nu}}{d\tau}\frac{dx^{\rho}}{d\tau}
\right.\nn
\left.+\accentset{(1)}{C}^{\mu}_{\nu\rho}\frac{dx^{\nu}}{d\tau}\frac{d\accentset{(1)}{\eta}^{\rho}}{d\tau}+\accentset{(1)}{C}^{\mu}_{\nu\rho}\frac{d\accentset{(1)}{\eta}^{\nu}}{d\tau}\frac{dx^{\rho}}{d\tau}+\partial_{\sigma}\accentset{(1)}C^{\mu}_{\nu\rho}\accentset{(1)}{\eta}^{\sigma}\frac{dx^{\nu}}{d\tau}\frac{dx^{\rho}}{d\tau}+\accentset{(2)}{C}^{\mu}_{\nu\rho}\frac{dx^{\nu}}{d\tau}\frac{dx^{\rho}}{d\tau}\right)
\end{gather}

\end{document}